\documentclass[useAMS,usenatbib]{mn2e}
\usepackage{graphicx}

\usepackage{subfigure}
\usepackage{amsmath}

\title[Bulgeless Discs II]{Hierarchical formation of bulgeless  galaxies II: Redistribution of angular momentum via  galactic fountains \\ }
\author[C. B. Brook et al.]{C. B. Brook$^{1}$, G. Stinson$^{1}$, B. K. Gibson$^{1,2}$, R. Ro\v{s}kar$^{3}$, J. Wadsley$^{4}$, T. Quinn$^{5}$\\
        $^1$Jeremiah Horrocks    Institute, 
            University of Central Lancashire, 
            Preston, PR1~2HE, UK \\
            $^2$Department of Astronomy \& Physics, Saint Mary's University, Halifax, Nova Scotia, B3H 3C3, Canada\\
            $^3$Institute for Theoretical Physics, University of Z\"{u}rich, Win- terthurerstrasse 190, CH-9057 Zurich, Switzerland\\
            $^3$Department of Physics and Astronomy, McMaster University, Hamilton, ON L8S 4M1, Canada\\
        $^5$Astronomy Department, University of Washington, 
            Box 351580, Seattle, WA 98195-1580, USA \\                  
                }

\begin{document}

\date{}

\pagerange{\pageref{firstpage}--\pageref{lastpage}} \pubyear{2011}

\maketitle

\label{firstpage}

\begin{abstract}
Within a fully cosmological hydrodynamical simulation, we form a  galaxy which rotates at 140 kms$^{-1}$,  and is characterised by two loose spiral arms and a bar, indicative of a Hubble Type SBc/d galaxy.   We show that our simulated galaxy  has no classical bulge,  with a pure disc profile at z=1, well after the major merging activity has ended. A long-lived bar subsequently forms, resulting in the formation of a secularly-formed ``pseudo" bulge,  with the final bulge-to-total light ratio B/T=0.21.   We show that the majority of gas which loses angular momentum and falls to the central region of the galaxy during the merging epoch is blown back into the hot halo, with  much of it returning later to form stars in the disc.   We propose that this mechanism of  redistribution of angular momentum via a galactic fountain, when coupled with the results from our previous study which showed why gas outflows are biased to have  low angular momentum, can solve the angular momentum/bulgeless disc problem of the cold dark matter paradigm. 

  \end{abstract}


\section{Introduction}

The cold dark matter paradigm successfully explains many properties of large scale structure observed in the Universe \citep{blumenthal84}. On galactic scales, however, some problems exist which cast doubt on the properties of dark matter, and the assumption that it is non-relativistic (cold), not the least of which is the ``angular momentum problem". 

The angular momentum of disc galaxies is  imparted by torques from large scale structure  on dark matter and gas at the time at which they begin to collapse under gravity \citep{peebles69,barnes87}. Disc galaxies will form, under the assumption that the angular momentum of the gas is conserved as it cools to the centre of the dark matter halo in which it resides \citep{fall80,white84}. Problems with this picture of disc galaxy angular momentum acquisition were highlighted in hydrodynamical cosmological simulations, which showed that when gas cools to the centre of dark matter halos and forms stars, subsequent merging of the resultant  proto-galaxies results in the catastrophic loss of angular momentum \citep{navarrowhite94}. This resulted in galaxies dominated by spheroidal components which were deficient in angular momentum when compared to their observed counterparts \citep{navarrosteinmetz00}. A closely related issue is that  bulges form during mergers, as gas  loses angular momentum and rapidly cools to form stars in  the centre of the galaxy \citep{barneshernquist96}.  These aspects of the angular momentum problem have significantly  improved with better modelling and resolution of  feedback processes, thus preventing stars from rapidly forming in the central regions  of the early collapsing dark matter halos \citep{thacker01,brook04,g07}. This is reflected in  recent simulations showing improved matches to the Tully-Fisher relation (\citealt{g07,agertz10,guedes11}, see also Section~\ref{properties} in this paper). 
  
Another manifestation of the problem is that the angular momentum {\it distribution} of simulated dark matter halos differs markedly from that of observed galaxies. This  is in conflict with the assumption that the distribution of the angular momentum of gas and dark matter are the same, an assumption which holds in non-radiative hydro-dynamical simulations \citep{vdB02,sharma05}. In particular,  dark matter halos have a significant amount of low angular momentum material \citep{bullock01}, and  in models of galaxy formation the corresponding low angular momentum baryons are expected to form a bulge component \citep{vdB01}. 
 
By contrast, many observed disc galaxies do not have bulge components (e.g. \citealt{allen06,kautsch09,cameron09}), and have angular momentum distributions that lack the significant amounts of  low angular momentum  that is found in dark matter halos \citep{vdBS01}.   The situation is worsened considerably when one considers that many bulges have characteristics that indicate that they were formed secularly by processes within the disc itself, rather than being related to merging or cold dark matter structure formation \citep{kormendy04,kormendy10}.

\begin{figure}%
\hspace{-.7cm} \includegraphics[height=.52\textheight]{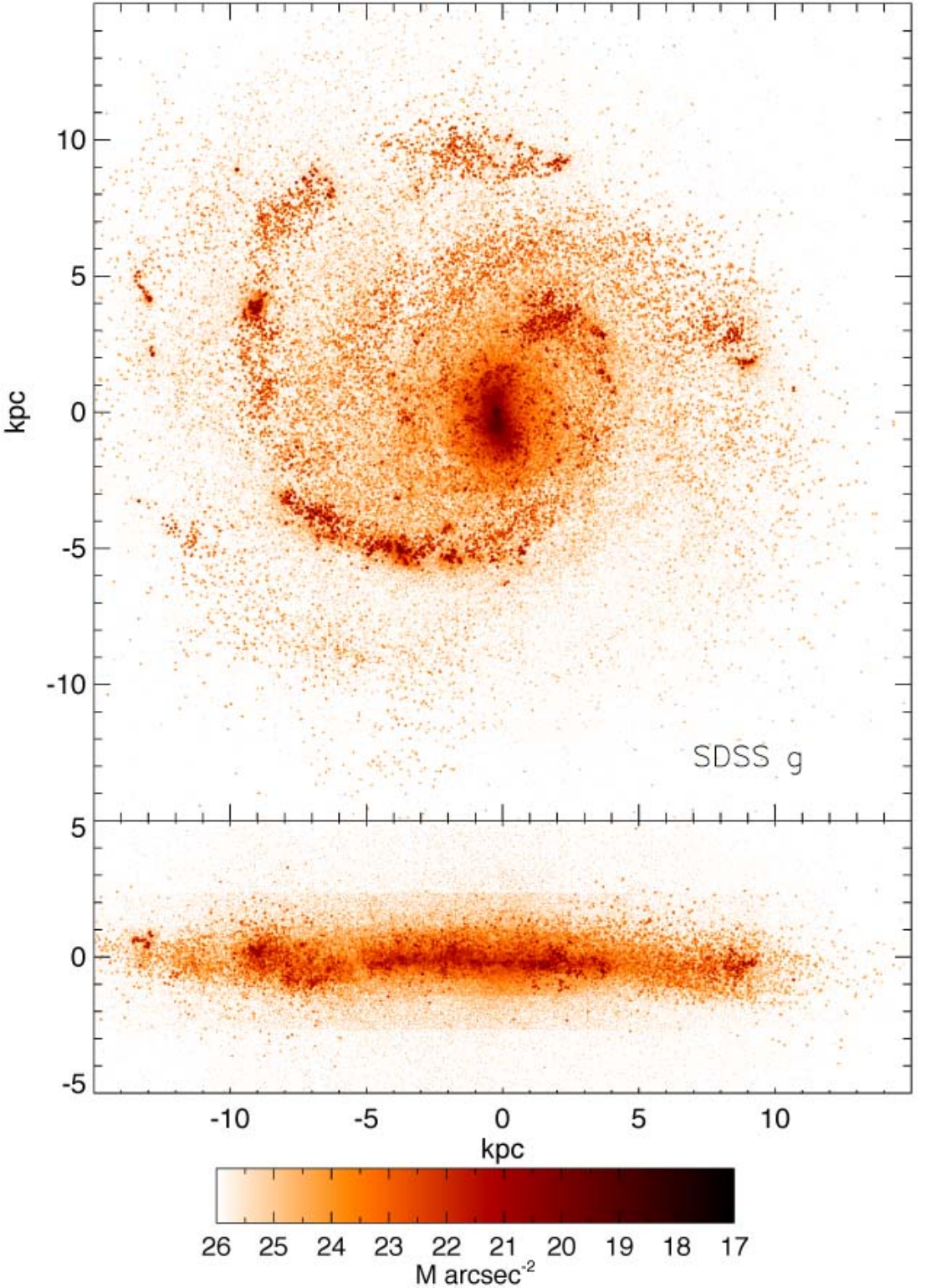}
  \caption{Face and edge on surface brightness maps in the SDSS g-band. Note the loose spiral arms and the existence of a bar. The effects of dust reprocessing have been included.}
\label{gband}
\end{figure}

The most recent simulations of massive disc galaxies continue to form classical  bulges, (B/T=0.19 in \citealt{agertz10}, and B/T=0.25 in \citealt{guedes11}, see also \citealt{scannapieco10,stinson10,brooks11}). We note that in the \cite{guedes11} paper, the authors claim that their bulge is  secular in nature, based on the low S\'{e}rsic index of the bulge in their two component bulge+disc decomposition. However, they also state that their B/T has doubled since $z=3$, indicating that their bulge is at least partially classical. 
These latest simulations have nevertheless formed significantly improved realisations of L$_*$ disc galaxies. Yet the angular momentum problem, and the associated problem of forming disc galaxies  which do not have  classical bulges, remains a serious challenge for the cold dark matter paradigm. 

Assuming  the verity of the cold dark matter paradigm, it would seem that  the resolution of the angular momentum problem requires either the {\it ejection} or {\it redistribution} of low angular momentum material. Ejection of low angular momentum material has certainly been  proposed  \citep{binney01,maller02,dutton09}. Indeed, the ejection of low angular momentum gas by supernova explosions was shown to be  key  to the  success of the first simulated  bulgeless dwarf galaxies, which share many characteristics with observed dwarf galaxies \citep{governato10,oh11}.  The processes which result in   the ejection of  low angular momentum gas  from low mass galaxies, and the key role this has  in suppressing bulge formation, were outlined in  Brook et al. (2011, Paper I).

The large scale  ejection of low angular momentum gas meshes well with the need to enrich the IGM \citep{maclow99,oppenheimer06,shen10}, and also with the observed low baryonic mass fraction  of low mass galaxies (e.g. \citealt{klypin99,mandelbaum06,koposov08,guo10,moster10}).  However, in the absence of AGN feedback, more massive galaxies  with larger potential wells will have greater difficulty ejecting their  gas. This is reflected in the strength of the relationship between total mass and the baryonic mass fraction;  higher mass disc galaxies have  significantly higher baryonic mass fractions, retaining  a far larger proportion  of  the cosmic baryon fraction. This means that the large scale outflow of low angular momentum gas is not likely to explain the fact that many massive disc galaxies do not have classical bulges \citep{kormendy10}.

\begin{figure}%
\hspace{-.3cm} \includegraphics[width=.35\textheight]{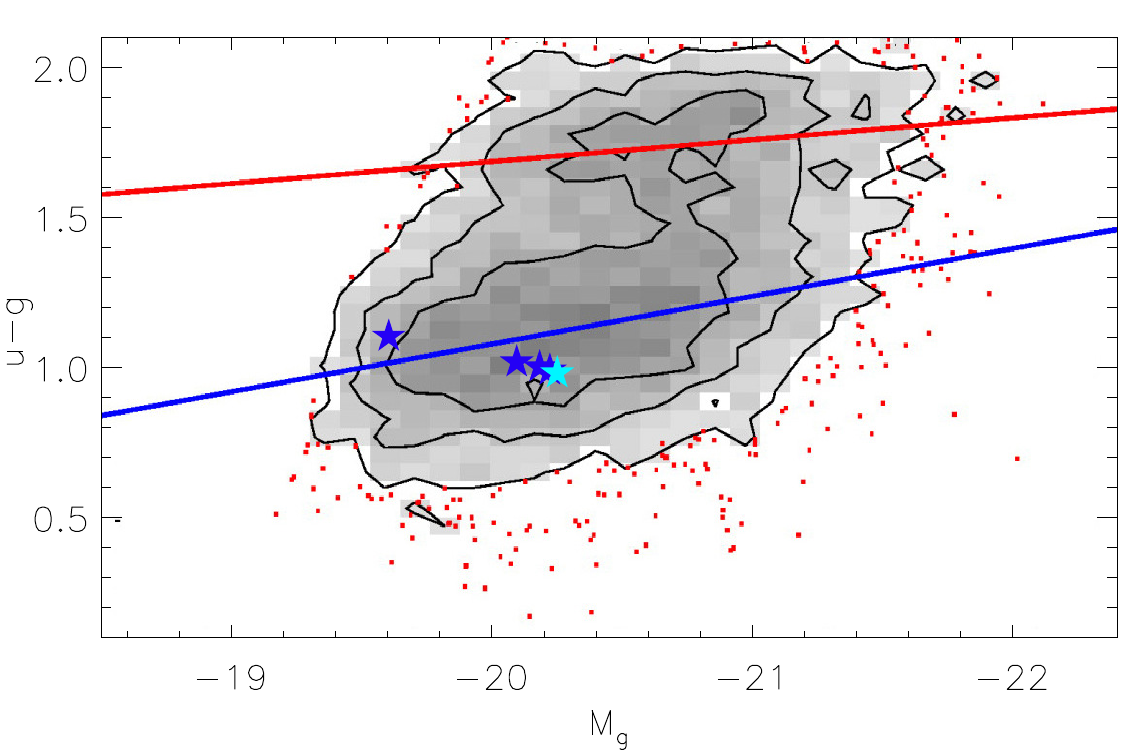}
  \caption{The M$_g$ $u$-$g$ colour magnitude diagram of SDSS galaxies from Blanton et al. 2006, with the trend of the blue and red clouds indicated. Overplotted as star symbols is our simulated galaxy. The light blue star is the face on value, while the dark blue stars show various angles of inclination ( 30$^{\circ}$, 45$^{\circ}$, 60$^{\circ}$ and 90$^{\circ}$ edge on, far left point). The effects of dust reprocessing are included, explaining the lower luminosity and redder colour of the simulation when viewed edge on.   }
\label{colmag}
\end{figure}

Thus, processes which  {\it redistribute} low angular momentum material may be necessary in higher mass disc galaxies. In this paper, we propose that galactic fountains are an effective mechanism for such redistribution. It is generally accepted that supernova-powered bubbles drive gas out of the galactic disc and through the halo. The expelled gas eventually falls back onto the disc, a mechanism known as a galactic fountain \citep{shapiro76, bregman80}.
Here, we examine galactic fountains on  larger scales, both spatially and temporally, by examining the fate of gas expelled from the inner regions of galaxies as they are assembling their mass at high redshift. 
It is well established that supernova-driven winds can result in gas outflows \citep{matthews71,veilleux05} and that  starbursts trigger strong  outflows from the central regions, particularly during merger events \citep{cc85,heckman90,strickland07,tremonti07}. Outflows are expected, and indeed observed to be more common at high redshift where star formation is more active, and potential wells shallower \citep{madau96,pettini98, pettini00,simcoe02, shapley03,adelberger05}. The  bipolar structure at the Galactic centre \citep{joss03,su10}  lends support for galactic scale outflows from the central regions, even locally.  
 The fate of such outflowing gas is not well constrained. As explained above, there is reason to believe that large amounts of the gas will be lost from low mass galaxies, but it is likely that some or even most of the outflowing gas   will cool back to the star forming region of higher mass galaxies.

\begin{figure}%
\hspace{-.5cm} \includegraphics[height=.23\textheight]{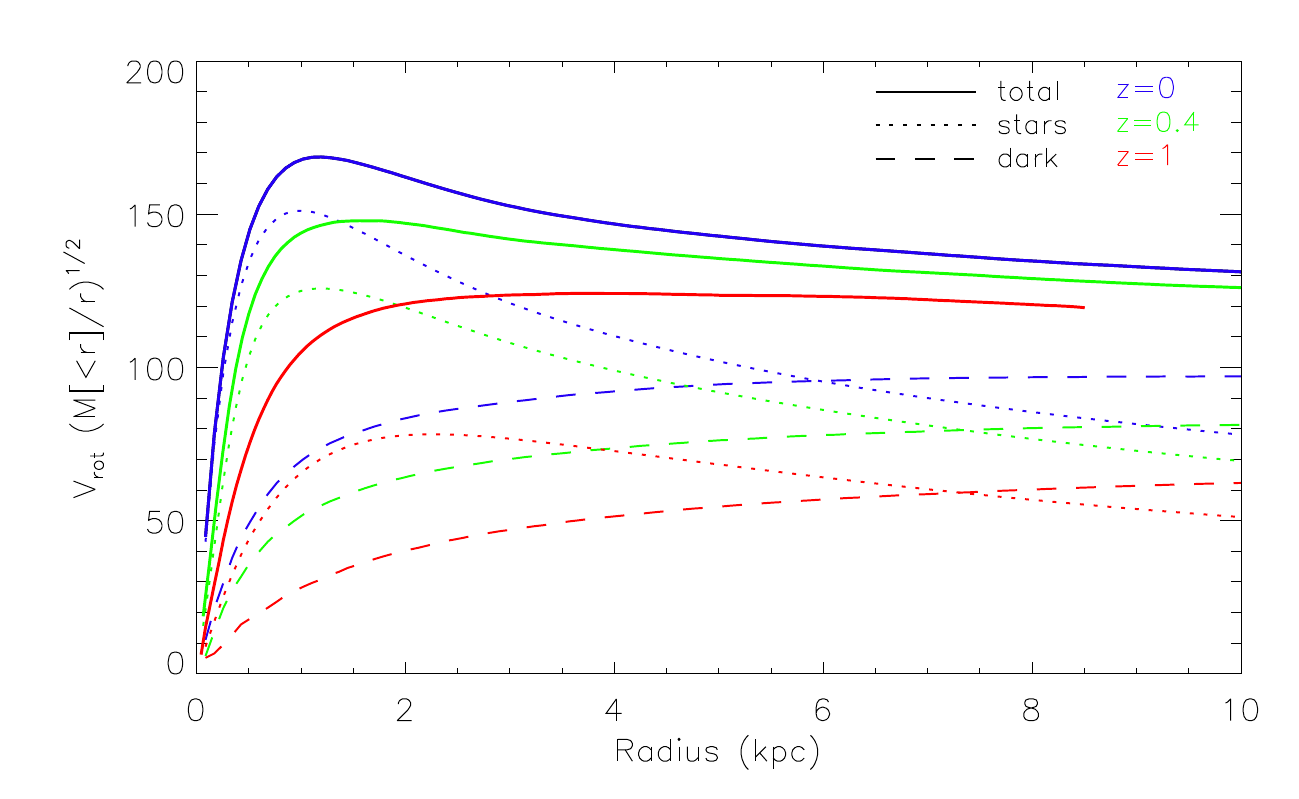}
  \caption{The rotation curve (V$_c=\sqrt{\frac{M(<r)}{r}}$) of the simulation at $z=0$ (blue line), $z=0.4$ (green) and $z=1$ (red). Also shown are the contributions from stars (dotted lines) and dark metter (dashed). The rotation curve is flat  at $z=1$, while it has a slight inner `peak' at $z=0$. }
\label{RC}
\end{figure}

\begin{figure}%
\hspace{-.5cm} \includegraphics[height=.31\textheight]{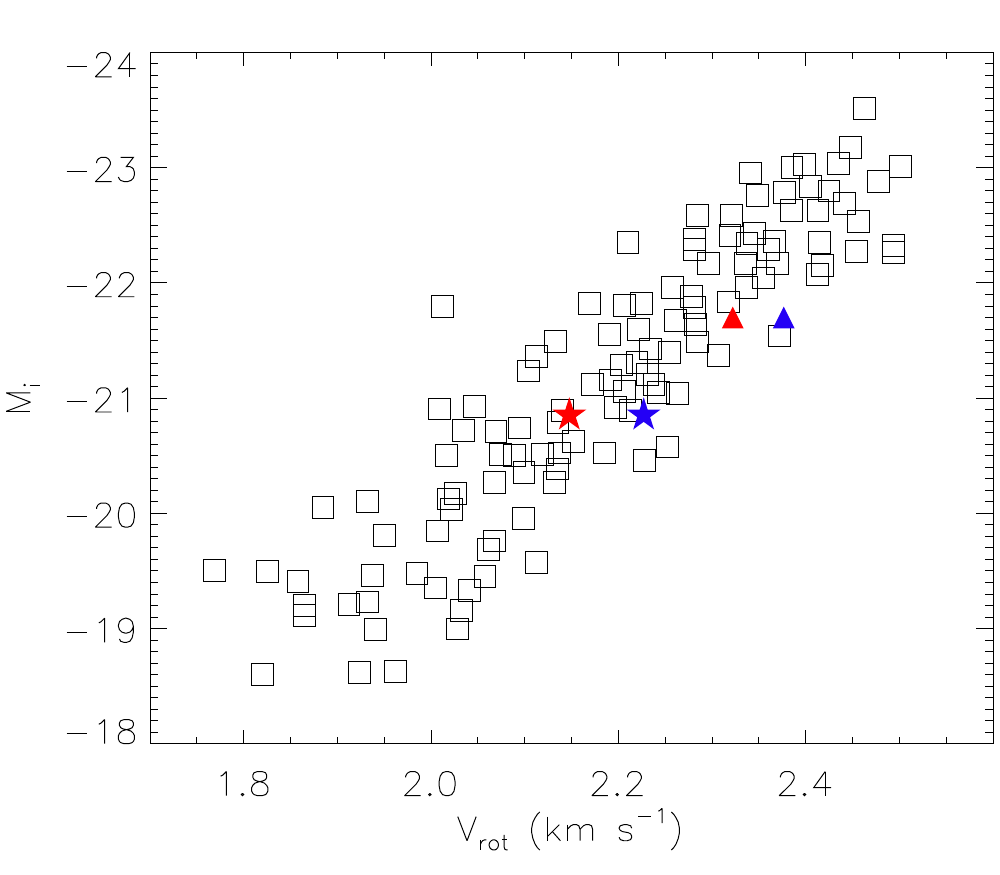}
 \caption{The  Tully-Fisher relation. Open squares: data  from Pizagno et al. 2007 using the SDSS i-band to measure V$_{80}$. Star symbols: our simulated galaxy, using  $V_{80}$ (red star)  and V$_{max}$ (blue star). Triangles: the recent ``Eris" simulation (Guedes et al. 2011), using V$_{80}$ (red triangle) and   V$_{max}$ (blue triangle).  }
\label{TF}
\end{figure}

Our model is based on analysis of our fully hydrodynamical cosmological galaxy formation simulations. Our simulation programme is running disc galaxies which span four orders of magnitude in stellar mass, from dwarf to $L_*$, all using the same resolution and input physics. We note that this input physics is similar to that which simulated the first bulgeless dwarf galaxies \citep{governato10}, and that we reproduce all the features of those successful simulations in our dwarfs. The simulation presented here also has similar input physics and resolution to the recent simulation of \cite{guedes11}, which formed a  much improved realisation of an L$_*$ galaxy. Here we examine a simulated disc galaxy  in an intermediate mass range, where the majority of observed  isolated galaxies are  dominated by their disc components. 

The paper proceeds by providing details of the simulation (Section~\ref{code}), and then describes the properties of the simulated galaxy (Section~\ref{properties}), and in particular we show that the simulated galaxy has no classical bulge. In Section~\ref{fountain}, we identify   gas which is blown from the bulge region and returns via the galactic fountain  to form stars at later times, and show that this re-accreted gas has  gained angular momentum before it forms stars. 
In Section \ref{discussion} we discuss our  modelling of   feedback from supernovae and its effect on the ISM,   central energy sources and angular momentum transport.
We conclude  (Section~\ref{summary}) that galactic fountains are effective in redistributing the angular momentum of gas, and propose that this mechanism plays a crucial role in reconciling observed massive disc galaxies which do  not have classical bulges with the $\Lambda$CDM paradigm.

\begin{figure*}%
\hspace{-.5cm}  \includegraphics[height=.23\textheight]{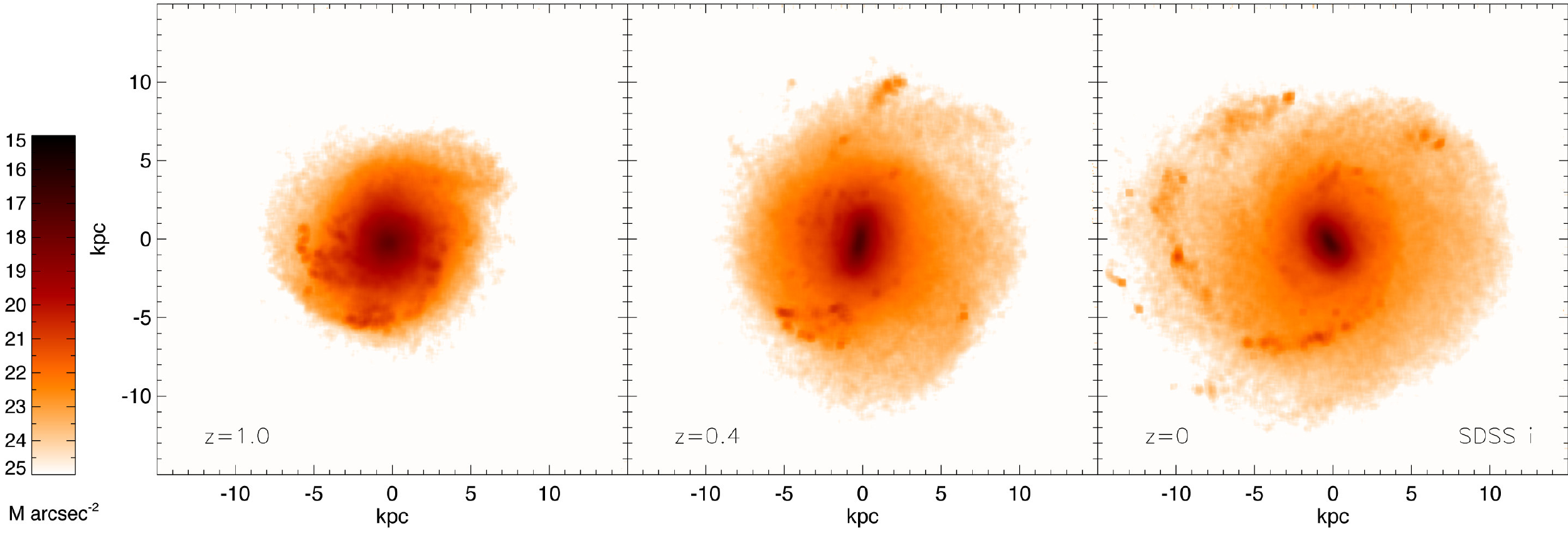}
  \caption{SDSS i-band surface brightness maps of the simulation at redshit $z=1$ (left), $z=0.4$ (middle) and $z=0$ (right). The disc grows inside out.A strong bar has developed by $z=0.4$   }
\label{Iband}
\end{figure*}

\section{The Simulation}\label{code}
The simulation described here is a cosmological zoom simulation derived from the McMaster Unbiased Galaxy Simulations (MUG
S).  See \citet{stinson10} for a complete description of how those galaxies were selected and their initial conditions were created.  Here, those initial conditions are scaled down, so that rather than residing in a 68 $h^{-1}$ Mpc cube, it is inside a cube with 34 $h^{-1}$ Mpc sides.  This resizing allows us to compare galaxies with exactly the same merger histories at a variety of masses. We selected a  MUGs galaxy  (15784)  which had a  prominent disk component for this study. The virial mass of the simulated galaxy in this study is $M_{vir}=19.4\times10^{10}$M$_{\odot}$

The simulation is evolved with the smoothed particle hydrodynamics (SPH) code \textsc{gasoline} \citep{wadsley04}.  \textsc{gasoline} employs  cooling due  to H, He, and a variety  of  metal lines, and is described in detail in \cite{shen10}.
The metal cooling
grid is constructed using CLOUDY (version 07.02,  \citealt{ferland98}), assuming ionisation equi-
librium. A uniform ultraviolet ionising background, \cite{haardtmadau96} is used in order to calculate the metal cooling rates
self-consistently. 

Two mechanisms are used to prevent gas from collapsing to higher densities than SPH can physically resolve: (i) to ensure that gas resolves the Jeans mass and does not  artificially fragment, pressure is added to the gas according to \citet{robertson08}, (ii) a maximum density limit is imposed by setting a minimum SPH smoothing length of 0.25 times that of   the gravitational softening length of $\epsilon=$ 155pc. We have $\sim$ 9 million resolution elements (1.1$\times10^{6}$ gas+5.3$\times10^{6}$ dark matter+2.3$\times10^{6}$ star) within the virial radius at $z=0$, with mean stellar particle mass of $\sim$ 6400~M$_{\odot}$.

\begin{figure}
\hspace{-.5cm}  \includegraphics[height=.3\textheight]{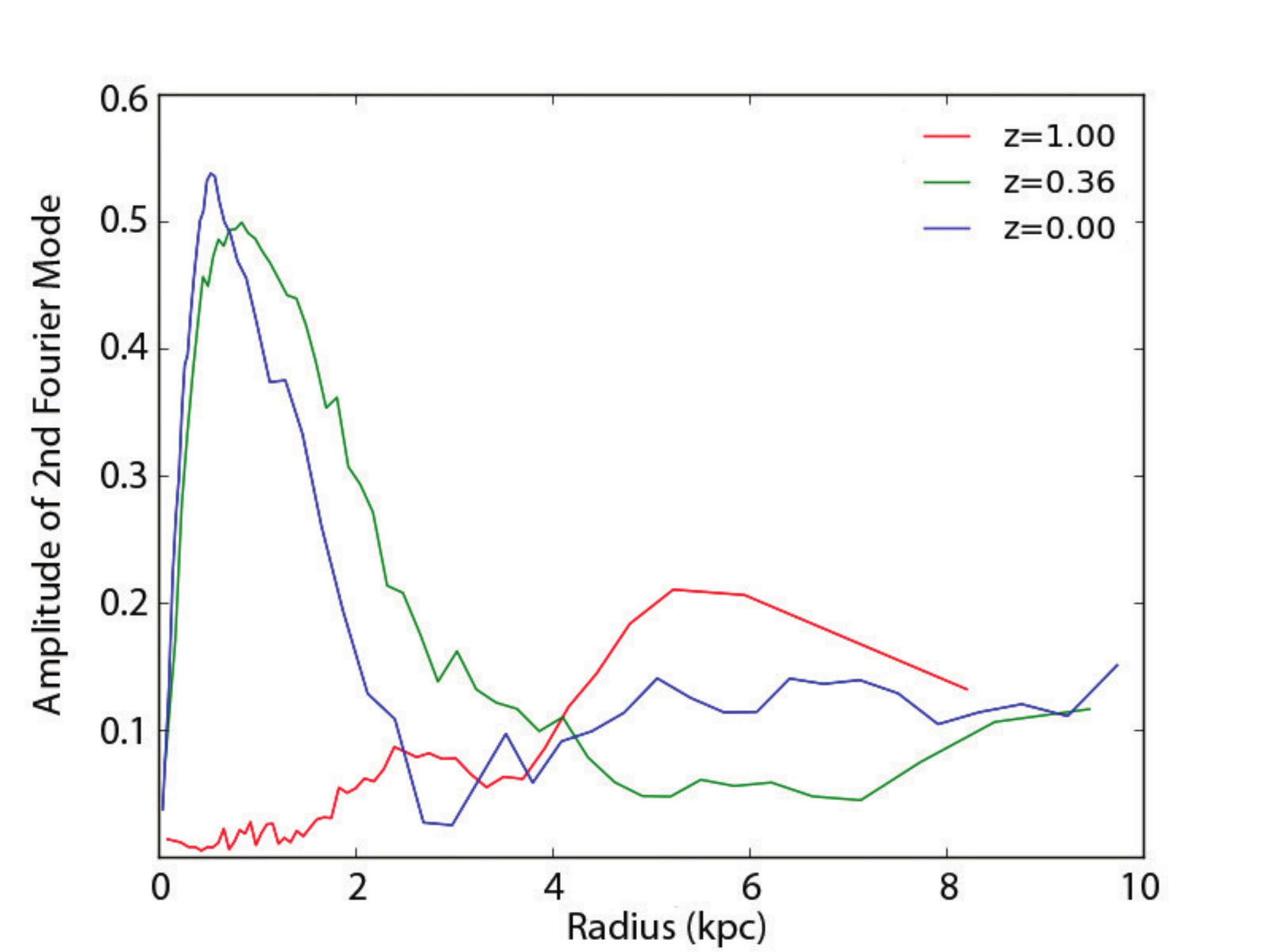}
  \caption{The amplitude of the 2$^{nd}$ Fourier mode of  the simulation at redshift $z=1$ (green), $z=0.4$ (red) and $z=0$ (blue).  The strong bar that develops by $z=0.4$ is not present at $z=1$.   }
\label{fourier}
\end{figure}

\subsection{Star Formation and Feedback}
When gas reaches cool  temperatures ($T < 15,000$ K) in a dense  environment ($n_{th} > 9.3$ cm$^{-3}$), it becomes eligible to form stars.  This value for $n_{th}$ is the maximum density gas can reach using gravity, $32M_{gas}/\epsilon{^3}$.  Such gas is converted to stars according to the equation
\begin{equation}
\frac{\Delta M_\star}{\Delta t} = c_\star \frac{M_{gas}}{t_{dyn}}
\end{equation}
Here, $\Delta M_\star$ is the mass of the star particle formed, $\Delta t$ is the timestep between star formation events, 0.8 Myrs in these simulations, $M_{gas}$ is the mass of the gas particle and $t_{dyn}$ is the gas particle's dynamical time.  $c_\star$ is the efficiency of star formation, in other words, the fraction of gas that will be converted into stars during $t_{dyn}$.

Stars feed energy back into the interstellar medium gas where they formed.  Two types of energetic feedback are considered in these simulations, supernovae and stellar radiation.  Supernova feedback is implemented using the blastwave formalism described in \citet{stinson06} and deposit $10^{51}$ erg of energy into the surrounding medium.  Since this gas is dense, the energy would be quickly radiated away due to the efficient cooling.  For this reason, cooling is disabled for particles inside the blast region $R = 10^{1.74}E_{\rm 51}^{0.32}n_0^{-0.16}\tilde{P}_{\rm 04}^{-0.20} {\rm pc}$ and for the length of time $t = 10^{6.85}E_{\rm 51}^{0.32}n_0^{0.34}\tilde{P}_{\rm 04}^{-0.70} {\rm yr}$ given in \cite{mckee77}. Here, $E_{\rm 51}=10^{51}$ erg, $n_0$ is the ambient hydrogen density, and
${P}_{\rm 04} = 10^{-4}P_0k^{-1}$ where $P_0$ is the ambient pressure and k is the
Boltzmann constant. Both $n_0$ and $P_0$ are calculated using the SPH
kernel for the gas particles surrounding the star.

Metals are ejected from type II supernovae (SNII), type Ia supernovae
(SNIa), and the stellar winds driven from asymptotic giant branch
(AGB) stars. ÊEjected mass and  metals are distributed to the nearest neighbour gas particles using the smoothing kernal. 
Details of the manner in which metals are incorporated into the code are found in  \citet{stinson06}. The prescribed SNII and SNIa yields have been updated in this study. ÊThe SNII
oxygen and iron use  fits to \citet{woosley95}. Ê
For SNIa, we use the W7 yields from Table~1 in \citet{nomoto97}. Metal diffusion is also included, such that unresolved turbulent mixing is treated
as a shear-dependent diffusion term, as described in \cite{shen10}. The allows proximate gas particles to mix their metals. Metal cooling is calculated
based on the diffused metals.

\begin{figure*}
\hspace{-.5cm}  \includegraphics[height=.18\textheight]{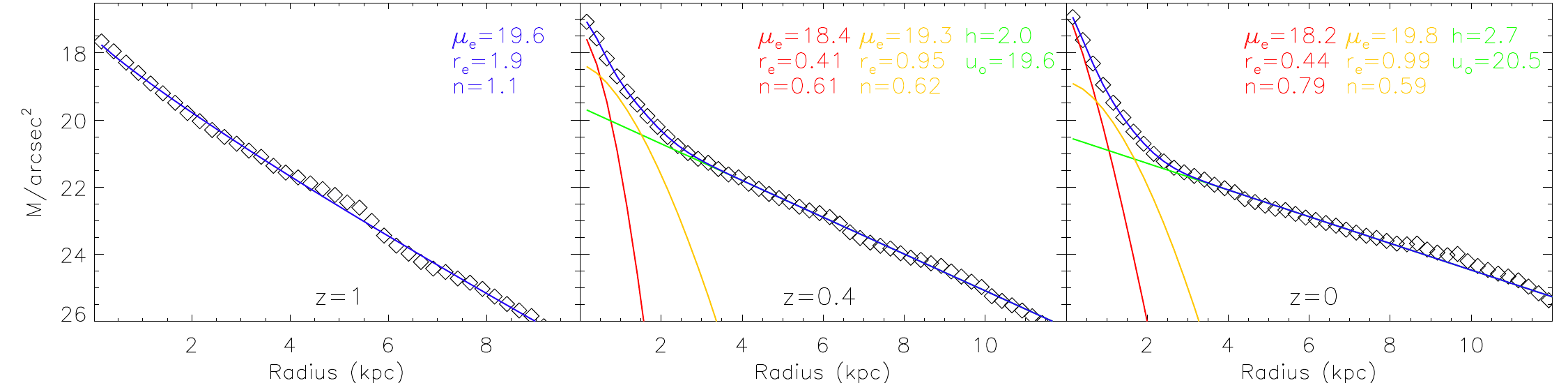}
  \caption{Decompositions of  the SDSS i-band surface brightness maps from Fig~\ref{Iband}. At $z=1$ (left panel), there is a pure disc galaxy, best fit by a single S\'{e}rsic fit having S\'{e}rsic index $n=1.1$. At $z=0.4$ (middle panel) and $z=0$ (right panel) we fit a bulge+bar+disc. We caution that the S\'{e}rsic indices of the bulge components are not reliable because of gravitational softening. We do not expect that the derived mass associated with each component (B/T $= 0, 0.12, 0.21$ at $z=1,0.4, 0$ respectively) will be affected by this.   }
\label{profiles}
\end{figure*}

The effects of stellar radiation have only recently begun to be considered in galaxy simulations \citep{hopkins11}.  Here, we model the luminosity of stars using the \citet{bressan93} mass-luminosity relationship. Typically, this relationship leads to $10^{53}$ ergs of energy being released from the high mass stars over the course of the 0.8Myr between star formation calculations.  These photons do not couple efficiently with the surrounding ISM  \citep{freyer06}.  We thus do not want to couple all of this energy to the surrounding gas in the simulation. To mimic this highly inefficient energy coupling, we inject 10\% of the energy as thermal energy in the surrounding gas, and cooling is \emph{not} turned off for this form of energy input. It is well established that such thermal energy injection is highly inefficient at the spatial and temporal resolution of  the type of cosmological simulations used here \citep{katz92}. This is primarily due to the characteristic cooling timescales in the star forming regions being lower than the step time of the simulations.    The dynamical effect of this feedback is negligible, as is the overall coupling of energy to the ISM, but high mass star radiation feedback does reduce star formation in the region immediately surrounding a recently formed star particle (full details of this radiative feedback will be presented in a separate paper, Stinson et al. in prep).

\section{The Simulated Galaxy: Properties}\label{properties}
The simulated galaxy has a total stellar mass of $1.4\times10^{10}M_{\odot}$, an HI mass of $1.9\times10^{9}M_{\odot}$ and a total gas mass (within R$_{vir}$) of $1.5\times10^{10}M_{\odot}$. The value $log(M(HI)/L_R)=-0.8$ fits right in the middle of the relation for observed  galaxies  Verheijen et al (2010). The disc has a metallicity which peaks at $[Fe/H]=-0.5$, while the cold gas has $O/H+12=8.25$.
The simulated galaxy's merger history is referred to  in the analysis which follows. The interaction of the  last major merger (mass ratio 2:1) begins at $z\sim2.7$, with coalescence finished by  $z\sim2.2$ and is followed rapidly by 2 other significant merger events (mass ratios of 10:1 and 20:1). We identify the period between $z=2.7$ and  $z=1.7$ as the galaxy's ``merger epoch".  Since $z\sim1.7$, several dark matter merger events occur, but these are too low in mass to contain baryons. 

In Figure~\ref{gband}, we plot the surface brightness profile at $z=0$ in the SDSS g-band, shown face-on and edge-on. The loose spiral arms are indicative of a late type galaxy. The existence of a bar is also evident, and we contend that the simulated galaxy would be classified as SBc/d. These surface profile maps were created with Sunrise  \citep{jonsson06}  which allows us to measure the
dust reprocessed spectral energy distribution (SED) of
every resolution element of our simulated galaxies from
the far UV to the far IR with a full 3D treatment of
radiative transfer. We assume a simple relationship between metallicity of gas, which is traced in the simulations,  and dust by using a constant dust-to metals ratio of 0.4 \cite{dwek98}. 
Filters mimicking those on major
telescopes are used to create mock observations. Sunrise uses
Monte Carlo techniques to calculate radiation transfer
through astronomical dust, including such effects in the
determination of SEDs.

The galaxy sits in the blue cloud in the  M$_g$, $u$-$g$ colour magnitude diagram of SDSS galaxies, reproduced from \cite{blanton06} in Figure~\ref{colmag}, with the trend lines of the blue and red clouds indicated. The simulated galaxy is shown  as star symbols; the light blue star shows the face-on values, while the dark blue stars show various angles of inclination (0$^{\circ}$, 30$^{\circ}$, 45$^{\circ}$, 60$^{\circ}$ and 90$^{\circ}$), with greater inclination resulting in redder colours and lower magnitudes. The inclusion of the  effects of dust reprocessing  explains the significantly  lower luminosity and redder colour of the simulation when seen edge-on.    

The rotation curve (V$_c=\sqrt{\frac{M(<r)}{r}}$) of the simulation is shown in Fig.~\ref{RC} at $z=0$ (blue line),  $z=0.4$ (green) and $z=1$ (red). Also shown are the contributions from stars (dotted lines) and dark matter (dashed) at those times. The flat curve at $z=1$ reflects the success of our simulation in preventing excessive star formation in the central regions during the merger epoch. The $z=0$ curve is not perfectly flat, and perhaps indicates that the secular processes which build the bulge may be too efficient, and also indicates that, despite significant improvements, our simulations do not yet exactly reproduce the mass distribution  of real galaxies. 

We plot the observed Tully-Fisher relation using data from \cite{pizagno07} who plot the SDSS i-band magnitudes  against V$_{80}$, the rotation velocity at the radius which encompass 80$\%$  of the i-band flux (open squares). Our simulated galaxy is overplotted using star symbols, where we show V$_{80}$ (red star)  to mimic the observations, as well as  V$_{max}$ (blue star), to show that the degree of the inner peak in our rotation curve is not ``catasptrophic". Also overplotted as triangles is the recent ``Eris" simulation \citep{guedes11}, with colours again indicating  V$_{80}$ (red triangle) and   V$_{max}$ (blue triangle).  Using similar inputs physics and resolution to our simulation, ``Eris" also has a good fit to  the Tully-Fisher relation which does  not depend on the radius at which the measurement is made.

The SDSS i-band surface brightness maps of the simulation at redshit $z=1$ (left), $z=0.4$ (middle) and $z=0$ (right) are shown in Fig~\ref{Iband}, with the surface brightness range indicated in the colour bar.  Note the inside out growth of the disc and the strong bar which has developed by $z=0.4$. The 2nd mode of the Fourier harmonics of the simulations is shown in Fig~\ref{fourier}, confirming that no bar exists at $z=1$, but a strong bar with $h_b/h_l\sim1.2$ has formed by $z=0.4$ (where $h_b$ is the bar length  and $h_l $ is the disc scale length). The bar weakens somewhat by $z=0$.

 In the three panels of Fig~\ref{profiles}, we show the  decompositions of the  SDSS i-band surface brightness maps from Fig~\ref{Iband}. Note that all these one dimensional profile  fits were cross-checked  GALFIT \citep{peng10}, producing similar results.  At $z=1$ (left panel), the simulation has formed  a pure disc galaxy, best fit by a single S\'{e}rsic profile having S\'{e}rsic index $n=1.1$. Recall that this is $\sim$2 Gyrs since the last significant merger event. At $z=0.4$ (middle panel) and $z=0$ (right panel) we fit a bulge+bar+disc. Also shown in the panels are the fit parameters, i.e. disc scale-length (h) and central surface brightness (u$_o$), and the three S\'{e}rsic parameters for the bulge and bar, i.e. the effective brightness (u$_e$), effective radius (r$_e$) and S\'{e}rsic index (n). In our decompositions, we derive very low S\'{e}rsic indices for the bulge component. We caution that this quantity is not reliable, due to gravitational softening which will ``expand" the inner mass distribution, with an effect out to $\sim 4$ softening lengths ($\sim 620pc$ in our simulation), which will lower the S\'{e}rsic index. The amount of light/mass in the derived component should not be affected, so the derived B/T    results are reliable (B/T$= 0., 0.12, 0.21$ at $z=1, 0.4, 0$ respectively).

Further, we note that other properties of secularly formed ``pseudo" bulges are not definitive \citep{kormendy04}, but more ``indicative". So, although our bulge is also flat ($\epsilon_{bulge}/\epsilon_{disc}$ ranges between 0.55 and 0.9 depending on viewing angle in relation to the bar, where $\epsilon$ is the elliptcity) and is an oblate rotator (V$_{max}/\sigma_V=1.4$, $\epsilon_{bulge} =0.48-0.75$, again depending on viewing angle),   by far the safest  evidence that our simulated galaxy's bulge formed secularly  is in the surface brightness profile evolution and its relation to merger events, from which we can determine that the bulge did not form from gas which lost its angular momentum due to interactions during the merger epoch, but rather it formed via secular processes during the life of the disc.

\section{Redistributing Angular Momentum}\label{fountain}
In this section, we identify and trace all gas which has lost angular momentum and cooled to the bulge region of the simulated galaxy during the galaxy's ``merging epoch". Specifically, we identify any cold gas  (T$<30,000$K)  within 2 kpc of the centre  of the galaxy at any time between $z=2.7$ and $z=1.7$. This gas generally leads to  classical bulge formation. In what follows we  refer to this gas as ``bulge gas" (BG). There is $3.0\times10^{9}M_{\odot}$ of such bulge gas, which is $\sim 20\%$ of the mass of stars in the galaxy at $z=0$. Note that time resolution in our output steps means that it is possible that we miss some of the cold gas which is in the bulge region during the merger epoch, so  $3.0\times10^{9}$M$_{\odot}$ is a lower bound. But there is no bias in our selection; the bulge gas that we trace, and from which our conclusions are drawn, is a representative sample of cold gas in the bulge region during the merger epoch.

\begin{figure}
\hspace{-.5cm} \includegraphics[height=.26\textheight]{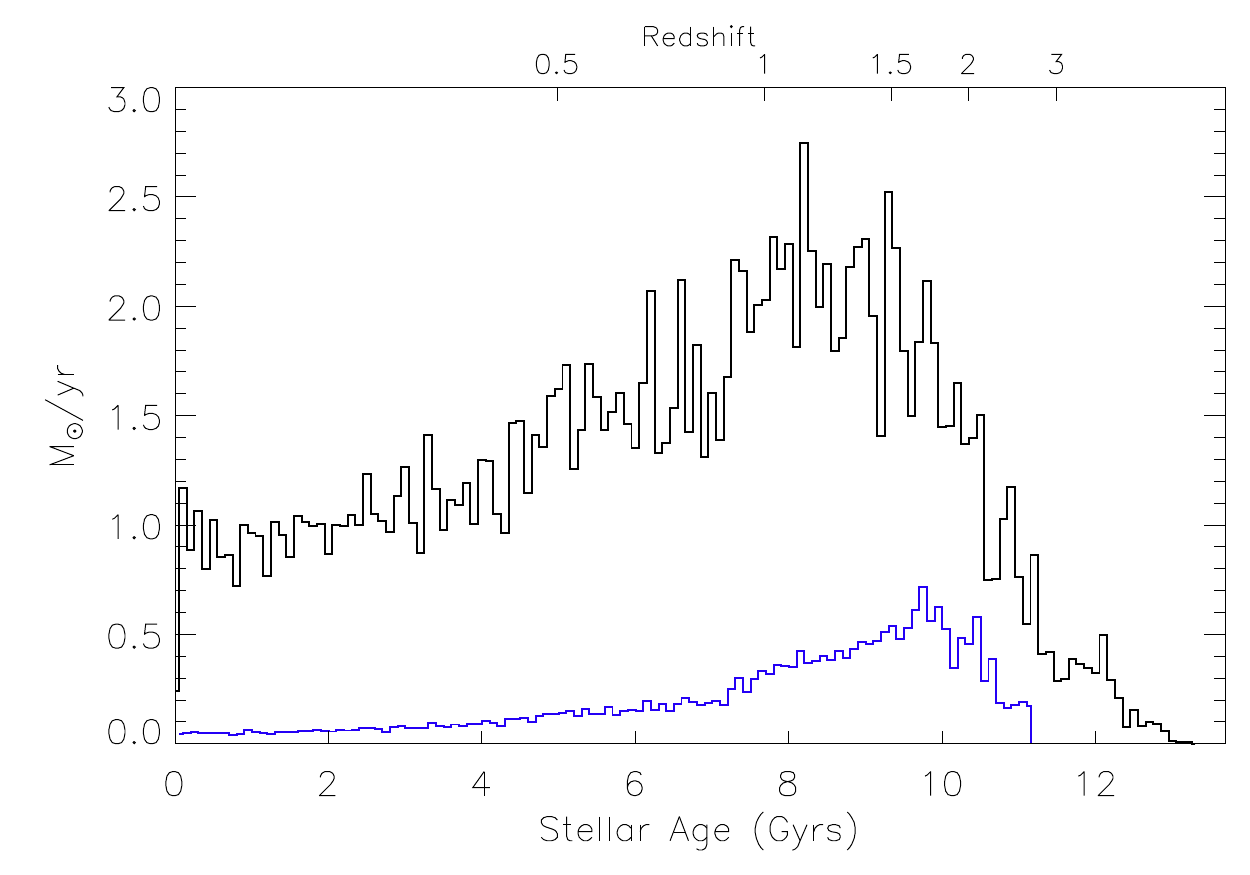}
  \caption{Black line: the star formation history of our simulation. Blue Line: the formation history of  stars which form from the bulge  gas  (BG), defined as any cold (T $<30,000$K) gas within 2 kpc of the centre of the galaxy during the merger epoch of the galaxy ($2.7>z>1.7$). 20\% of all stars from from  BG. Redshifts are shown on the upper x-axis.   }
\label{sfr}
\end{figure}

By $z=0$, 71\% ($2.13\times10^{9}M_{\odot}$)  of the bulge gas has formed stars.  Of the  29\% ($8.7\times10^{8}M_{\odot}$) remaining as gas at $z=0$, 14\% ($1.2\times10^{8}M_{\odot}$)  has been blown entirely out of the galaxy, 75\%  ($6.5\times10^{8}M_{\odot}$) is in the hot halo, while 11\% ($1.0\times10^{8}M_{\odot}$) has subsequently cooled back down to form HI  and is  feeding star formation in the disc. 
In Fig~\ref{sfr} we plot the star formation history of all  stars in the simulated galaxy at $z=0$ (black line). The blue line shows  all stars which form from the bulge gas.   Redshifts are included on the upper x-axis. Note the extended star formation history of the stars formed from bulge gas. Although 71\% of bulge gas forms stars by $z=0$, only 9\% of it forms stars rapidly after it was first identified as being  in the bulge during the merger epoch. The star burst from the 9\% bulge gas which forms stars during the merger epoch ejects the  remaining 91\% of the bulge gas  from the bulge region, much of which returns to the galaxy  and forms stars at a later  time. We note that these large scale outflows  also act to prevent gas from falling  to the central region during this merger epoch, enhancing their effectiveness at preventing bulge formation.

\begin{figure}
\hspace{-.5cm} \includegraphics[height=.26\textheight]{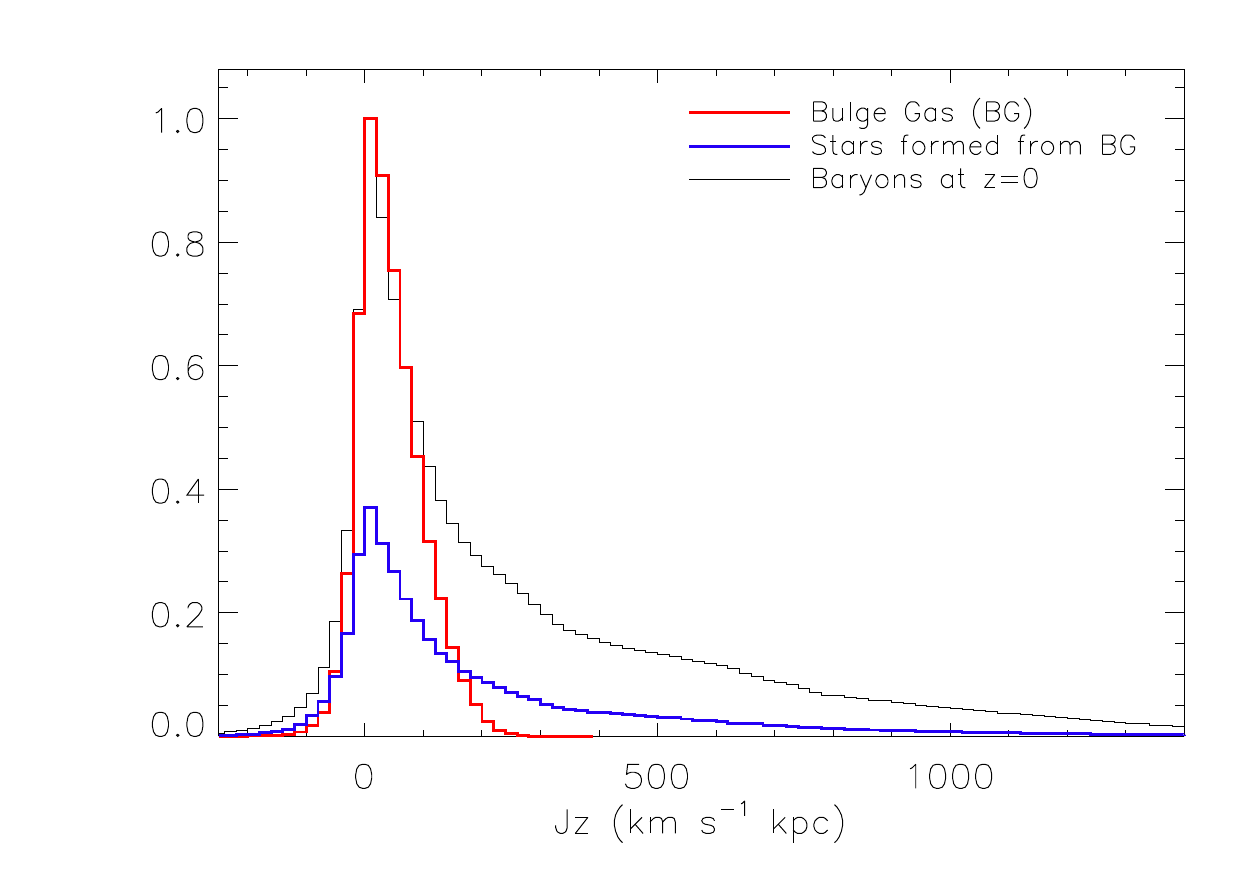}
  \caption{The angular momentum distribution of  bulge gas (BG, red line) and stars  at $z=0$ which form from the BG (blue line). These lines are normalised so that the area under the curves indicates the mass. Also shown  is the  angular momentum distribution of observable ``baryons" at $z=0$, where baryons = stars + HI gas (black line), normalised to have a peak at the same height as the bulge gas. The shape of the distribution of stars formed from BG (blue line) closely resembles the $z=0$ baryon distribution, while the BG (red line) has  low angular momentum.  
  }
\label{jzhist}
\end{figure}

\begin{figure}
\hspace{-.2cm} \includegraphics[height=.65\textheight]{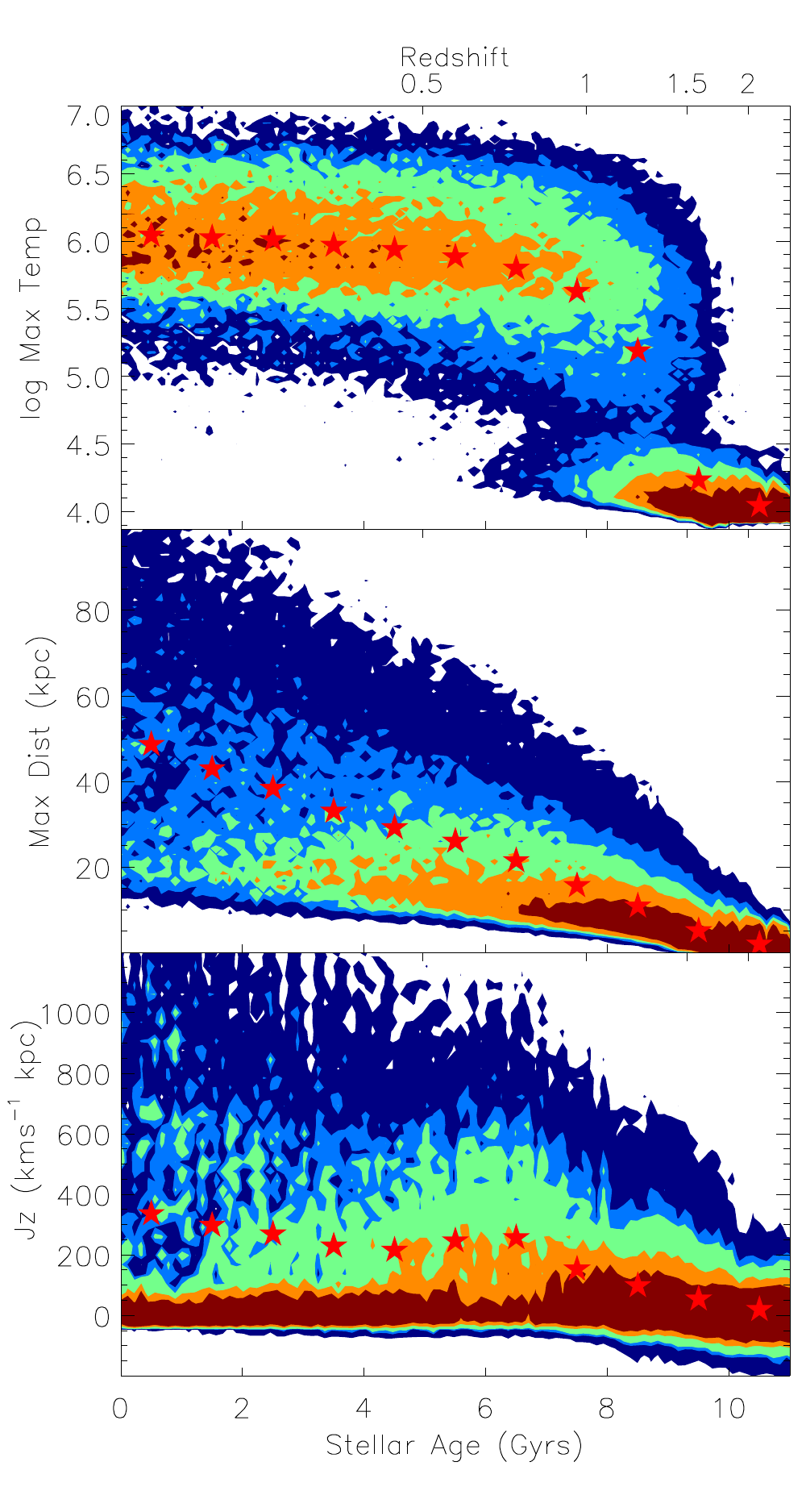}
  \caption{Top panel: the mean of the log of the Maximum Temperature that   ``star forming bulge gas" (BGs, see text)  reaches subsequent to the time that it was identified, plotted against the age of the stars which form from that BGs.    Middle panel: the mean of the maximum distance  that   star forming bulge gas (BGs)  reaches from the centre of the galaxy, plotted against the age of BGs stars. Bottom panel:  the mean angular momentum of BGs stars  is plotted against their age. Red star symbols show the median values at each age. Supernova feedback heats gas and ejects it from the central region with the hottest gas travelling the furthest from the centre. The ejected gas absorbs angular momentum from the hot corona and later accreting gas, returning to the disk with a significantly different angular momentum distribution.
 }
\label{trendstime}
\end{figure}

Figure~\ref{jzhist} shows the angular momentum distribution of bulge gas (BG, red line) at the time that it was identified. The blue line shows the  $z=0$ angular momentum distribution of the stars which form from the BG. These two lines are normalised so that the area under the curves indicates the mass. Also shown for perspective (black line) is the  angular momentum distribution of observable ``baryons" at $z=0$, where baryons = stars + HI gas.
 The baryons distribution is normalised to have a peak at the same hight as the bulge gas. The median (mean) values for the BG, stars formed from BG, and $z=0$ baryons are  35.7 (43.9), 92 (205) and 164 (305) kms$^{-1}$ kpc respectively.The shape of the distribution of stars formed from BG (blue line) closely resembles the $z=0$ baryon distribution, while the BG (red line) has universally low angular momentum. The angular momentum of the bulge gas has been dramatically {\it redistributed} during the process of being blown out of the central region and cooling back to the disc via  the galactic fountain. Rather than form bulge stars, the re-accreted   gas now primarily forms disc stars.

In Fig.~\ref{trendstime}, we trace the history of the 71\% of the  bulge gas which forms stars, and refer to this as ``star forming bulge gas" (BGs). Redshifts are shown on the upper x-axis. In the top panel, the log of the Maximum Temperature that  BGs  reaches subsequent to the time that it was identified (i.e. after they were measured as being in the inner 2 kpc with T $<30,000$K) is plotted against the age of the stars which form from BGs.  In the middle panel, the maximum distance  that   BGs  reaches from the centre of the galaxy subsequent to the time that they were identified, is plotted against the age of BGs stars. In the bottom panel,  the mean angular momentum of BGs stars  is plotted against their age. 9\% of the BG forms stars in the bulge region shortly after the time that  it is first identified, before it is blown out of the central region, and consequently forms these stars with the low angular momentum it had when identifed (Fig~\ref{jzhist}). Supernova feedback heats gas which has low angular momentum and removes it from the inner central region of the galaxy, with the hottest gas travelling the furthest from the centre. The ejected gas mixes with the corona gas, as well as later accreting gas, and  returns to the disk with a significantly altered angular momentum distribution. In particular, gas that is  re-accreted from the galactic fountain at later times has a significanlty less negative and low angular momentum baryons, a far greater amount of high angular momentum material, a higher median angular momentum, and forms stars in the disc rather than bulge. 

\section{Discussion }\label{discussion}
\subsection{Transition from Ejection to Redistribution}
  In Paper I and in this study, we have shown that in our simulations,  there are two ways in which galaxies can alter the angular momentum distribution of their baryons: {\it ejection} in large scale outflows and {\it redstribution} of low angular momentum material in galactic fountains. Ejection was shown to be a crucial process in simulated dwarf galaxies, and this meshes well with the low baryonic mass fractions, while redistribution was shown to be a crucial process in the more massive simulated galaxy considered in this study. Both processes occur at both mass scales in our simulations. Lower than universal baryon fractions  indicating that {\it ejection} may play a central role at all galaxy mass scales. 
   
  We caution that the  relative significance of  {\it ejection} to {\it redistribution}   is probably not precisely reproduced in current  simulations.  This is due to the uncertainty    involving the inclusion of feedback and the manner in which it couples to the ISM, a problem which is inherent to  all galaxy formation models.  In a semi-analytic study which assumes the ejection of low angular momentum gas,  \cite{dutton09} found that it was necessary to couple 25\% of SNe energy to outflows to produce realistic disc galaxy populations. Current   estimates of the relationship between  stellar mass and halo mass \citep{mandelbaum06,koposov08,guo10,moster10,dutton10,more11,avilareese11} indicate that the current simulation has a factor $\sim2-3$ too many stars, implying that the role of  {\it ejection} of low angular momentum gas  was most likely {\it underestimated} for a galaxy of the mass considered in this study. We note that the key observable for detecting ejection  is in fact baryon  rather than stellar mass fraction, but the fact that our simulation has a reasonable value of HI/L$_R$ means that we can assume that we have a similar over-abundance of baryons for our halo mass.  
  What we do expect is the relative importance of {\it redstribution} compared to {\it ejection}  to increase in more massive disk galaxies, compared to their low mass counterparts, and that the {\it ejection} fraction due to SNe  would be the lowest for L$_*$ galaxies where the ratio of stellar mass \citep{guo10,moster10}   and baryonic mass \citep{bell03} to dark halo mass are both at their highest. 
  More massive systems, even groups and clusters of galaxies,  also have a baryon fraction that is less than universal \citep{mccarthy07,giodini09,dai10} but this is widely attributed to the affects of Active Galactic Nuclei which are not included in these simulations.

 \subsection{Additional  Central Energy Sources}

  Our simulations do not include all possible energy sources, and any extra central heating, for example from black holes, nuclear star clusters \citep{mclaughlin06,rodriguez11} or a top heavy IMF  in the bulge \citep{ballero07}, would add to the efficacy of our model. We again note that the outflows due to  starbursts triggered during  merger events actually {\it prevent} gas flowing to the central regions, enhancing their effectiveness (it was shown in Paper I that it is preferentially low angular momentum gas above and below the plane that is prevented from being accreted).

  \subsection{Angular Momentum Transfer}\label{AM}
In a previous study of angular momentum transfer in similar galaxy formation simulations, \cite{roskar10},  showed that outer disc warps result from torquing of the cold gas  by a massive (misaligned) hot gas halo. Due to this torquing in the outer halo, the cold gas aligns with the spin of the hot gas halo by the time it reaches the central region. In those simulations, all the  gas (clumpy, shocked, cold flow, recycled from disc) ended up with  the same angular momentum distribution, due to this torquing,  coupled with  hydrodynamical  drag. 
These processes are what is also occurring in our simulations.

On the scale of local galactic fountains, significantly higher  resolution simulations can be employed than can be achieved in fully cosmological studies such as ours.  Hydrodynamical simulations of these local galactic fountains have shown that both drag  and  accretion into clouds can efficiently transfer angular momentum between the galactic corona and gas outflows \citep{fraternali08,marinacci11}.
Although there are many differences between the temperatures, times scales and  velocities of the interactions of the galactic fountains in those studies and in ours,   this is encouraging. Certainly, there is a need for significant  theoretical and observation work to constrain the transport of angular momentum between  gas with different origins and phases, particularly in the corona where hot halo gas mixes with newly accreted gas and gas blown from the star forming regions of the galaxy.

  We also note that disc viscosity and gas pressure within the disc due to supernovae and stellar winds \citep{valenzuela07} may play a role in angular momentum transport, but some angular momentum must be imparted to the gas prior to re-accretion for it to fall at large enough radii to form part of the disc.

\section{Summary}\label{summary}
 We have formed a simulated, late type  disc galaxy which has no classical bulge. The gas which is driven to the bulge region during the merger epoch, and thus has low angular momentum, is blown out  by supernova powered winds which follow  the star bursts that are triggered by the mergers. A significant fraction of the blown out  gas is re-accreted at later times and forms disc stars. We show that this redistribution of angular momentum via large scale galactic fountains can lead to the  formation of  massive disc galaxies which do not have classical bulges. In Paper I, we showed that galactic outflows will result in the {\it ejection} of low angular momentum material and hence the formation of bulgeless dwarf galaxies. The processes outlined in Paper I which relate to {\it ejection} of gas remain relevant to biases toward low angular momentum outflows from more massive galaxies, whether or not such gas is re-accreted, and the two papers should be read in conjunction.

 We have shown that our self-consistent model  for the {\it ejection} and {\it redistribution} of low angular momentum gas, with the latter increasing in  significance in higher mass galaxies,  can explain the existence of bulgeless disc galaxies that form hierarchically within a cold dark matter universe.  Our model has the implicit assumptions that (i) energy from supernova during merger triggered starbursts couples effectively to the ISM  and (ii) angular momentum can be  transported between  between ejected gas and newly accreted and hot halo gas. These processes require significant theoretical and observational work to be better constrained, which will provide crucial tests of our model.

\section*{Acknowledgments}

CB and BKG acknowledge the support of the UKÕs Science \& Technology Facilities Council (STFC Grant ST/F002432/1). TQ was supported by NSF ITR grant PHY- 0205413. We acknowledge the computational support provided by the UKÕs National Cosmology Supercomputer, COSMOS.
\bibliographystyle{mn2e}
\bibliography{brook}

\end{document}